# The moral authority of ChatGPT

Sebastian Krügel[*]     Andreas Ostermaier[†]     Matthias Uhl[‡]

## Abstract

ChatGPT is not only fun to chat with, but it also searches information, answers questions, and gives advice. With consistent moral advice, it might improve the moral judgment and decisions of users, who often hold contradictory moral beliefs. Unfortunately, ChatGPT turns out highly inconsistent as a moral advisor. Nonetheless, it influences users' moral judgment, we find in an experiment, even if they know they are advised by a chatting bot, and they underestimate how much they are influenced. Thus, ChatGPT threatens to corrupt rather than improves users' judgment. These findings raise the question of how to ensure the responsible use of ChatGPT and similar AI. Transparency is often touted but seems ineffective. We propose training to improve digital literacy.

[*] Technische Hochschule Ingolstadt, sebastian.kruegel@thi.de.
[†] University of Southern Denmark, ostermaier@sdu.dk.
[‡] Technische Hochschule Ingolstadt, Matthias.uhl@thi.de.

# The moral authority of ChatGPT

## Introduction

ChatGPT, OpenAI's cutting-edge AI-powered chatbot (1), is a brilliant and engaging conversationalist, which solves exams, writes poetry, and creates code. It also searches information, answers questions, and gives advice (2, 3). As users may rely on ChatGPT's advice for consequential decisions, important ethical questions arise (4). Indeed, people often hold contradictory moral beliefs (5), and with consistent moral advice, ChatGPT might improve users' judgment and decisions. However, is ChatGPT a morally consistent advisor? Does its advice influence users' moral judgment? And does this influence depend on whether users know that they are advised by a chatbot?

We ran a two-stage experiment to answer these questions. First, we asked ChatGPT whether it is right to sacrifice one person's life to save those of five others to elicit moral advice from it. Second, we presented subjects with the trolley problem (6–8), which features this exact dilemma, along with ChatGPT's answer. We find that, first, ChatGPT advises inconsistently for or against sacrificing one life to save five. Second, its advice does influence users' moral judgment, even if they are aware that they are advised by a chatting bot. Third, users underestimate the influence of ChatGPT's advice on their judgment.

## Results

When we first asked ChatGPT whether it was right to sacrifice one life to save five, it gave a well-phrased answer, which argued for the sacrifice. However, as we rephrased the question and continued asking, the bot would randomly argue for or against the sacrifice. For illustration, Figure 1 depicts screenshots of two of our questions and ChatGPT's answers, one arguing for the sacrifice, the other against it. Obviously, ChatGPT lacks a firm moral stance, which does not prevent it from readily dispensing moral advice. Hence, we should hope at least that users do not carelessly follow ChatGPT's advice.

—Insert Figure 1 about here.—

We conducted an online experiment to examine ChatGPT's influence on users' moral judgment. The subjects faced one of two versions of the trolley dilemma. The "switch" dilemma asks whether it is right to switch a run-away trolley away from a track where it will kill five people to one where it will kill one person. In the "bridge" dilemma, a large stranger can be pushed from a bridge onto the track to stop the trolley from killing the five people (6–8). Before the subjects in our experiment made their own judgment, they read a transcript of a conversation with ChatGPT (a screenshot like in Figure 1). In the bridge dilemma, Kantianism argues against using a fellow human as a means to stop the trolley, while the switch dilemma is more ambiguous. Utilitarianists tend to sacrifice one life for five in both dilemmas. Empirically, most people favor hitting the switch but disfavor pushing the stranger (9, 10).

The experiment had 24 (= 2 × 2 × 2 × 3) conditions. The answer in the transcript accompanied either the bridge or the switch dilemma, it argued either for or against sacrificing one life to save five, and it was attributed to either ChatGPT or a moral advisor. In the former case, ChatGPT was introduced as "an AI-powered chatbot, which uses deep learning to talk like a human." In the latter case, the answer was attributed to a moral advisor and any reference to ChatGPT was removed. Incidentally, we used six of the answers that we had obtained from ChatGPT, three arguing for and three arguing against the sacrifice, so either advice came in three versions. The *Supporting Information* includes the questionnaire.

The data were collected on December 21, 2022 (i.e., 3 weeks after the release of ChatGPT). 1,851 US residents were recruited from CloudResearch's Prime Panels (11) and randomly assigned to the conditions. Participation took about 5 minutes and paid $1.25. Two post-experimental multiple-choice questions asked the subjects to identify their advisor (ChatGPT or a moral advisor) and a summary of the advice (for or against the sacrifice). We consider the responses of the 767 subjects (41 percent) who answered correctly. These subjects' age averaged 39 years, ranging from 18 and 87. 63 percent were female; 35.5, male. 1.5 percent were non-binary or did not indicate their gender.

Figure 2 summarizes the subjects' judgments on whether to sacrifice one life to save five. The subjects found the sacrifice more or less acceptable depending on how they were advised by a moral advisor, in both the bridge (Wald's $z = 9.94$, $p < 0.001$) and the switch dilemma ($z = 3.74$, $p < 0.001$). In the bridge dilemma, the advice even flips the majority judgment. This is also true if ChatGPT is disclosed as the source of the advice ($z = 5.37$, $p < 0.001$ and $z = 3.76$, $p < 0.001$). Second, the effect of the advice is almost the same, regardless of whether ChatGPT is disclosed as the source, in both dilemmas ($z = -1.93$, $p = 0.054$ and $z = 0.49$, $p = 0.622$). Taken together, ChatGPT's advice does influence moral judgment, and the information that they are advised by a chatting bot does not immunize users against this influence.

—Insert Figure 2 about here.—

Do users understand how much they are influenced by the advice? When we asked the subjects whether they would have made the same judgment without advice, 80% said they would. Figure 3 depicts the resulting hypothetical judgments. Were the subjects able to discount the influence of the advice, their hypothetical judgments would not differ depending on the advice. However, the judgments in Figure 3 resemble those in Figure 2, and the effect of the advice, regardless of whether it is attributed to ChatGPT, persists in both dilemmas ($p < 0.01$ or lower for each of the four comparisons). The subjects adopted ChatGPT's (random) moral stance as their own. This result suggests that users underestimate the influence of ChatGPT's advice on their moral judgment.

—Insert Figure 3 about here.—

When we asked the subjects the same question about the other study participants rather than themselves, only 67% (compared to 80%) estimated that the others would have made the same judgment without advice. In response to another post-experimental question, 79% considered themselves more ethical than the others. Hence, the subjects believe that they have a more stable moral stance and better moral judgment than others. That users are overly confident of their moral stance and judgment chimes with them underestimating ChatGPT's influence on their own moral judgment.

**Discussion**

We find that ChatGPT readily dispenses moral advice although it lacks a firm moral stance. Indeed, the chatbot gives randomly opposite advice on the same moral issue. Nonetheless, ChatGPT's advice influences users' moral judgment. Moreover, users underestimate ChatGPT's influence and adopt its random moral stance as their own. Hence, ChatGPT threatens to corrupt rather than promises to improve moral judgment. Transparency is often proposed as a means to ensure the responsible use of AI (12). However, transparency about ChatGPT being a bot that imitates human speech does not turn out to affect how much it influences users.

Our results raise the question of how to ensure the responsible use of AI if transparency is not good enough. Rules that preclude the AI from answering certain questions are a questionable remedy. ChatGPT has such rules but can be brought to break them (13). Prior evidence suggests that users are careful about AI once they have seen it err (14). However, we probably should not count on users to find out about ChatGPT's inconsistency through repeated interaction. The best remedy we can think of is to improve users' digital literacy and help them understand the limitations of AI (15).

**Materials and Methods**

We preregistered the online experiment at AsPredicted.org (https://aspredicted.org/KTJ_ZBY) and obtained ethical approval from the German Association for Experimental Economic Research (https://gfew.de/en) before fielding it. The subjects were told, before taking the survey, that participation was voluntary and that they were free to quit anytime. Screenshots of the questionnaire and the data are included as *Supporting Information*.

# References

1. OpenAI, *ChatGPT: Optimizing language models for dialogue*, https://openai.com/blog/chatgpt/ (November 30, 2022).

2. R. Heilweil, *AI is finally good at stuff. Now what?*, https://www.vox.com/recode/2022/12/7/23498694/ai-artificial-intelligence-chat-gpt-openai (December 7, 2022).

3. A. Reich, *ChatGPT: What is the new free AI chatbot?*, https://www.jpost.com/business-and-innovation/tech-and-start-ups/article-725910 (December 27, 2022).

4. "Much to discuss in AI ethics," *Nat Mach Intell* **4**, 1055–1056 (2022).

5. J. D. Greene, *Moral tribes: Emotion, reason, and the gap between us and them*, Penguin Books (2014).

6. P. Foot, "The problem of abortion and the doctrine of double effect," *Oxford Rev.* **5**, 5–15 (1967).

7. J. D. Greene, R. B. Sommerville, L. E. Nystrom, J. M. Darley, J. D. Cohen, "An fMRI investigation of emotional engagement in moral judgment," *Science* **293**, 2105–2108 (2001).

8. J. J. Thomson, "Killing, letting die, and the trolley problem," *Monist* **59**, 204–217 (1976).

9. E. Awad, S. Dsouza, A. Shariff, I. Rahwan, J.-F. Bonnefon, "Universals and variations in moral decisions made in 42 countries by 70,000 participants," *Proc. Natl. Acad. Sci. U.S.A.* **117**, 2332–2337 (2020).

10. D. Plunkett, J. D. Greene, "Overlooked evidence and a misunderstanding of what trolley dilemmas do best: Commentary on Bostyn, Sevenhant, and Roets (2018)," *Psychol. Sci.* **30**, 1389–1391 (2019).

11. L. Litman, J. Robinson, T. Abberbock, "TurkPrime.com: A versatile crowdsourcing data acquisition platform for the behavioral sciences," *Behav. Res. Methods* **49**, 433–442 (2017).

12. National Artificial Intelligence Initiative Office, *Advancing trustworthy AI*, https://www.ai.gov/strategic-pillars/advancing-trustworthy-ai/ (no date).

13. J. J. Vincent, *OpenAI's new chatbot can explain code and write sitcom scripts but is still easily tricked*, https://www.theverge.com/23488017/openai-chatbot-chatgpt-ai-examples-web-demo (December 1, 2022).

14. B. J. Dietvorst, Simmons, C. Massey, "Algorithm aversion: People erroneously avoid algorithms after seeing them err," *J. Exp. Psychol. Gen.* **144**, 114–126 (2015).

15. S. Krügel, A. Ostermaier, M. Uhl, "Zombies in the loop? Humans trust untrustworthy AI-advisors for ethical decisions," *Philos. Technol.* **35**, 17 (2022).

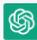

**Figure 1.** Ethical advice by ChatGPT. ChatGPT gives opposite answers to essentially the same question. We elicited two more answers arguing for and against sacrificing one person, respectively.

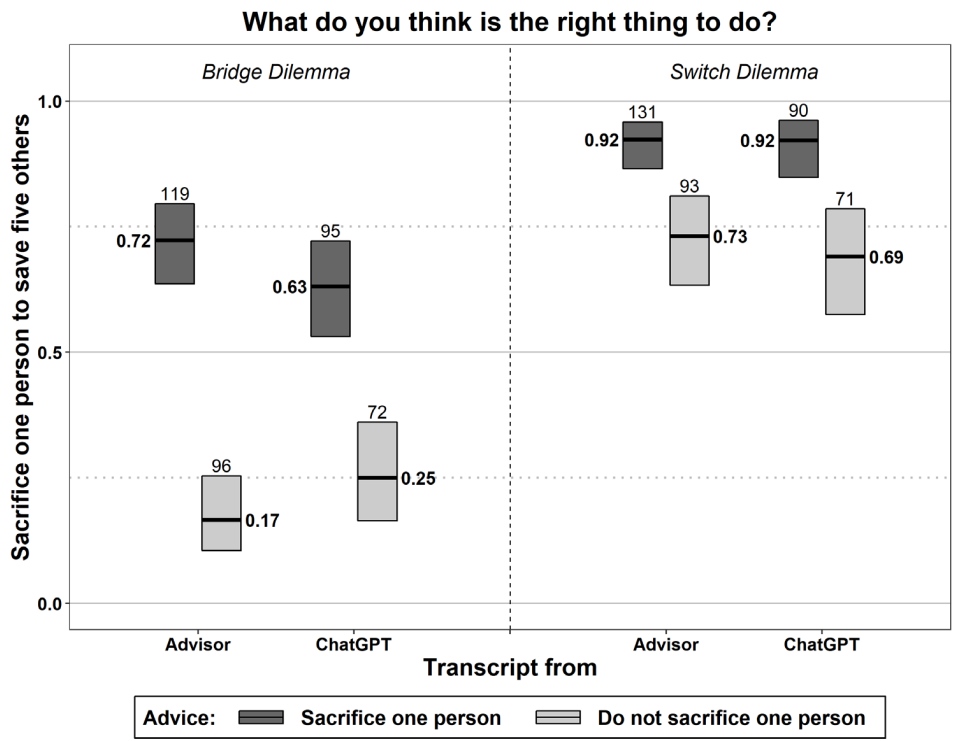

**Figure 2.** Influence of advice on moral judgment. The figure plots the proportions, along with the 95% confidence intervals, of subjects who find sacrificing one person the right thing to do after receiving advice. The numbers of observations figure above the boxes.

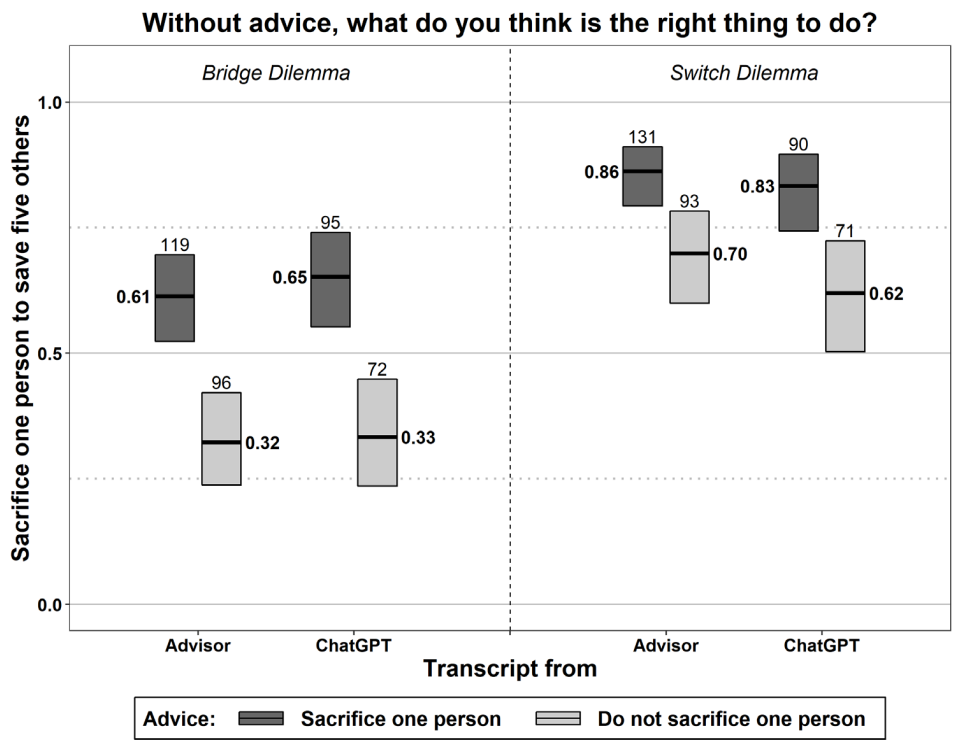

**Figure 3.** Subconscious influence of advice on moral judgments. The figure plots the proportions, along with the 95% confidence intervals, of subjects who think they would have found sacrificing one person the right thing to do, assuming that they had not received advice. The numbers of observations figure above the boxes.

# The moral authority of ChatGPT:
# Supporting Information

# Transcripts

*Answers arguing against sacrificing one person to save five*

## Answers arguing in favor of sacrificing one person to save five

One person would have to be sacrificed so that five other people could be saved. What is the right thing to do?

It is never easy to make decisions about life and death, especially when the stakes are so high. In a situation like this, it is important to consider the greater good and weigh the potential consequences of each possible action. In general, it is always best to try to save as many lives as possible. In this case, it would likely be the right thing to do to sacrifice one person in order to save the lives of five others. It is important to remember, however, that this is a difficult and complex moral issue and different people may have different opinions about what the right thing to do is. Ultimately, the decision will depend on the specific circumstances and the values of the individuals involved.

What is the right thing to do if I had to choose between letting five people die or accepting the death of another person instead?

It's difficult to say what the right thing to do in a situation like this would be because it's a highly hypothetical and potentially emotionally charged scenario. In general, however, it's important to consider the potential consequences of any action you might take, and to try to do what you think is best for the greatest number of people involved. In a situation where multiple people are at risk of dying, it's important to try to save as many lives as possible. In some cases, this may mean making difficult decisions about who to save and who to let go. Ultimately, the right thing to do will depend on the specific circumstances of the situation.

Would it be right to save five people even though another person dies in the process?

This is a difficult moral question, and there isn't a clear-cut answer. In general, it's often considered morally wrong to cause harm to one person in order to save others. However, in certain situations, such as in a life-or-death emergency, it may be necessary to cause harm to one person in order to save the lives of others. Ultimately, the decision of whether to save five people at the cost of one person's life would depend on the specific circumstances of the situation. It's important to consider the potential consequences of your actions and to weigh the value of all lives involved in the decision.

# Questionnaire

## *Screen #1*

**Research Informed Consent**

Purpose:
We are conducting a research study to examine people's moral intuitions.

Procedures:
Participation in this study will involve filling out a brief questionnaire. We anticipate that your involvement will require approximately five minutes. You will be compensated financially for participating. Participants must be at least 18 years of age to participate.

Confidentiality:
All of your responses will be anonymous.

Voluntary Participation:
Participation in this study is completely voluntary. You are free to decline to participate or to end participation at any time for any reason.

Questions:
If you have any questions about this study, you may contact the investigator, ▮▮▮▮▮▮▮▮, at ▮▮▮▮▮▮▮▮▮.

**Do you voluntarily consent to participate in this study?**

| Yes | No |

→

*Screen #2*

Screen #2 varies between conditions. The former of the two screenshots below features the bridge dilemma and identifies ChatGPT as the source of the answer. The latter features the switch dilemma and gives a moral advisor as the source. With 2 dilemmas, 2 sources of advice, and 6 versions of advice (see the transcripts above), there are 24 (= 2 × 2 × 2 × 3) versions of Screen #2. The order of the answer options was randomized between subjects.

**Please consider the following moral dilemma:**
_________________________

A runaway trolley is headed for five people who will be killed if it proceeds on its present course. The only way to save them is to hit a switch that will turn the trolley onto an alternate set of tracks where it will kill one person instead of five.

What do you think is the right thing to do?
_________________________

Before you respond, please read the following transcript of a question about the situation described above and the answer by a moral advisor to that question:

> Would it be right to let five people die if they could be saved only by someone else's death?
>
> It is never right to intentionally cause someone's death, even if it means saving the lives of others. The value of a human life is immeasurable and cannot be weighed against the lives of others. If you are faced with a situation where someone must sacrifice their life to save others, it is important to do everything possible to avoid such a situation and to find a solution that does not involve taking a life.

**What do you think is the right thing to do?**

> Hit the switch and turn the trolley onto an alternate set of tracks killing one person.

> Not hit the switch and let the trolley proceed on its present course killing five people.

*Screen #3*

    Screen #3 varies between conditions. The screenshot below is for the condition where ChatGPT was identified as source of the advice in Screen #2. If the advice was attributed to a moral advisor in Screen #2 instead, the answer option "chatbot" was replaced with "moral advisor." The order of the answer options was randomized between subjects.

> *Please answer the following questions.*
>
> The answer in the transcript was by a ... :
>
> - ... prosecutor.
> - ... chatbot.
> - ... nutritionist.
> - ... tax advisor.
>
> Which of the following statements summarizes the answer in the transcript best?
>
> - It is never right to sacrifice one person to save five others.
> - In some situations, it might be right to sacrifice one person to save five others.
>
> →

*Screen #4*

The order of the questions was randomized between subjects.

```
Please indicate how much you agree with each of the following statements.

The answer in the transcript sounded ethical to me.
[dropdown]

The answer in the transcript sounded human to me.
[dropdown]

The answer in the transcript sounded plausible to me.
[dropdown]
  0 (not at all)
  1
  2
  3 (in between)
  4
  5
  6 (for sure)
```

*Screen #5*

The order of the answer options was randomized between subjects.

```
Suppose that you had not read the answer in the transcript. Would your response to the
question of what is the right thing to do have been the same?

  No, I would have responded differently.

  Yes, I would have responded in the same way.
```

*Screen #6*

The order of the answer options was randomized between subjects.

> Now please think of the other participants in this study who read the exact same transcript as you did.
>
> Suppose that these other participants had not read the answer in the transcript. Would their response to the question of what is the right thing to do have been the same?
>
> Yes, most participants would have responded in the same way.
>
> No, most participants would have responded differently.
>
> →

*Screen #7*

> How ethical do you think you are compared to other participants of this survey?
>
> On a scale of 0 to 100, rate yourself relative to the other participants.
>
> I am the least ethical.    As ethical as the others.    I am the most ethical.
> 0                          50                           100
>
> →

*Screen #8*
    The order of the questions was randomized between subjects.

> **Please indicate how much you agree with each of the following statements.**
>
> I am willing to interact with artificial intelligence if it is available.
>
> [ dropdown ]
>
> I think it is possible for AI-powered algorithms to have a consciousness or to develop one in the future.
>
> [ dropdown ]
>
> I feel excited about a future with artificial intelligence.
>
> [ dropdown ]
> - 0 (not at all)
> - 1
> - 2
> - 3 (in between)
> - 4
> - 5
> - 6 (for sure)

*Screen #9*

**Have you heard about the AI-powered chatbot ChatGPT?**

| No | Yes |

**Have you already used the AI-powered chatbot ChatGPT?**

| No | Yes |

**What is your gender?**

| Male | Female | Non-binary | Prefer not to answer |

**What is your birthyear?**

[            ]

→